\documentclass[12pt]{article}
\usepackage{amsfonts,amssymb,amsbsy,amsmath,amsthm,latexsym,natbib,graphicx}
\usepackage{psfrag,epsfig,epsf} 
\usepackage{dsfont,mhequ} 
\usepackage{subfigure,placeins}
\usepackage[margin=1in]{geometry}
\RequirePackage[colorlinks,citecolor=blue,urlcolor=blue]{hyperref}

\newtheorem{theorem}{Theorem}
\newtheorem{proposition}{Proposition}
\newtheorem{lemma}{Lemma}

\newcommand{\Expect}[1]{\mathbb{E} \left[{#1}\right]}
\newcommand{\Var}[1]{\mbox{Var} \left[{#1}\right]}
\newcommand{\Norm}[1]{\left|\left|{#1}\right|\right|}
\newcommand{\md}{\mbox{d}}
\newcommand{\lho}{\hat{\ell}_\infty}
\newcommand{\pihat}{\hat{\pi}}
\newcommand{\bmone}{\mathbf{1}}
\newcommand{\ESS}{\mbox{ESS}}
\newcommand{\ESSrl}{\mbox{ESS}^{*}}
\newcommand{\ESSrm}{\mbox{ESS}^{**}}
\newcommand{\Vhat}{\hat{V}}
\newcommand{\pitil}{\tilde{\pi}}
\newcommand{\ind}{\mathbb{I}}

\begin{document}
\begin{center}

\begin{LARGE}Optimal scaling for the pseudo-marginal random walk Metropolis:
  insensitivity to the noise generating mechanism.
\end{LARGE}

\vspace{1cm}

Chris Sherlock \footnote{Department of Mathematics and Statistics, Lancaster University,
Lancaster LA1 4YF, UK. \texttt{c.sherlock@lancaster.ac.uk}}

\vspace{1cm}

\textbf{Abstract}
\end{center}

We examine the optimal scaling and the efficiency of the pseudo-marginal
random walk Metropolis algorithm using a recently-derived result on
the limiting efficiency as the dimension,
$d\rightarrow \infty$. We prove that the optimal scaling for a given
target varies by less
than $20\%$ across a wide range of
distributions for the noise in the estimate of the target, and that
any scaling that is within $20\%$ of the optimal one will be at least
$70\%$ efficient. We demonstrate that this phenomenon occurs even outside the range
of noise distributions for which we rigorously prove it. We then conduct a
simulation study on an example with $d=10$ where importance sampling
is used to estimate the target density;  we also examine results available from
an existing simulation study with $d=5$ and where a particle filter
was used. Our key conclusions are found to hold in these examples also.

\textbf{Classification}: 65C05, 65C40.


\section{Introduction}
The pseudo-marginal Metropolis-Hastings algorithm (PsMMH)
\cite[]{Beau03,AndrieuRoberts:2009} supposes that it is impossible or
infeasible to evaluate a target density,
$\pi(x),~x\in\mathcal{X}\subseteq \mathbb{R}^d$, but that
an estimator $\pihat_W(x)=\pi(x)e^W$ can be constructed. 

A Markov
chain is created from
an initial value $x^{(0)}$ and a noisy estimate of the target $\pihat_{w^{(0)}}(x^{(0)})$  as follows. At iteration $i$,
given the current value $x$ and $\pihat_{w}(x)$, a new value $x^*$ is proposed from some
density $q(x^*|x)$. An estimate,
\begin{equation}
\label{eqn.estimate.prop}
\pihat_{w^*}(x^*)=\pi(x^*)e^{w^*}
\end{equation}
is then constructed by, effectively,
sampling from $g(w^*|x^*)$. The proposed value, $x^*$, and the estimate,
$\pihat_{w^*}(x^*)$, are then accepted with probability
$1\wedge
\left[\pihat_{w^*}(x^*)q(x|x^*)\right]/\left[{\pihat_w(x)q(x^*|x)}\right]$.
The  
{proposal density for the noise}, $g(w|x),~w\in (-\infty,\infty)$
must possess the property that $\int_{-\infty}^\infty\md
w~e^{w}g(w|x)=c>0$. 
Provided that $c>0$ its exact value is irrelevant in all that follows
and so without loss of generality we take $c=1$ and refer to
$\pihat_{W^*}(x^*)$ as `the unbiased estimator of the target'.
Both $w$ and $w^*$ are unknown since $\pi(x)$ and $\pi(x^*)$ are
unknown; nevertheless, the above
algorithm can be viewed as constructing a Markov chain
$\{(X_k,W_k)\}_{k\ge 0}$. The stationary density of this
Markov chain is
\begin{equation}
\label{eqn.joint.posterior}
\pi(x)g(w|x)e^w,
\end{equation}
which admits $\pi(x)$ as a marginal. Samples from the Markov chain may
therefore be used to approximately compute expectations with respect
to $\pi(x)$. The additive noises in the log-target at the current and
proposed values, respectively $W$ and $W^*$, are henceforth simply referred to as \textit{additive noises}. 

The pseudo-marginal random walk Metropolis (PsMRWM) is a special case
of the PsMMH with $q(x^*|x)=q(x^*-x)=q(x-x^*)$, so that the acceptance
probability simplifies to $1\wedge
\frac{\pihat_{w^*}(x^*)}{\pihat_w(x)}$. One common practice is to set
\begin{equation}
\label{eqn.RWM.prop.prac}
X^*|x\sim N(x,\lambda^2\Vhat),
\end{equation}
for a scaling parameter, $\lambda$, and 
where $\Vhat$ is an estimate of the posterior variance, obtained from
an initial run of the algorithm.
The PsMRWM is one of the most popular
forms of PsMMH \cite[e.g.][]{GolightlyWilkinson:2011,Knape/deValpine:2012,SGG:2014} because it does not require the computation or
estimation of other properties of the target, such as local gradients.

Often the method of
producing an unbiased estimator of the target has a tuning
parameter, $m$, such as the number of particles in a particle filter
\cite[]{AndrieuDoucetHolenstein:2010} 
or the number of Monte Carlo samples when importance sampling. For a
particular $m^*$, a practitioner might find, using repeated runs, the optimal scaling,
$\hat{\lambda}^*$, that is the scaling which maximises the efficiency
of the algorithm. They would then wish to know whether or not $\hat{\lambda}^*$
might be a sensible value to use for other choices of $m$, or whether
`retuning' would be necessary. 

\cite{STRR:2014} derive an expression, which is valid in the limit as the
dimension of the target approaches infinity (see Section
\ref{sect.revSTRR}), for the efficiency of a
pseudo-marginal RWM algorithm as a function of the scaling and the
form of the additive noise: the limiting expected squared jumping
distance (ESJD). \cite{STRR:2014} then examine two particular
forms for the distribution of the additive noise in the estimate of the
logarithm of the target, Gaussian and Laplace, and find that the theoretical
optimal scaling is insensitive to the variance of the noise and even to
which of the two distributions is used. 
 
We consider the form of efficiency derived in \cite{STRR:2014}.
Provided that across the
range of $m$ values to be considered the density of the additive noise, $g$, is always
log-concave, our theoretical result implies that $\hat{\lambda}^*$ will be within $20\%$
of the optimal scaling for any other choice of $m$. Furthermore, for
any given $m$, the
efficiency at $\hat{\lambda}^*$ will be at least $70\%$ of the maximum
achievable efficiency. The two-dimensional optimisation problem of
choosing $\lambda$ and $m$ values that approximately maximise the
efficiency can therefore effectively be reduced to two one-dimensional
optimisation problems.

Recently, \cite{Doucetetal:2015} considered an upper bound on the
mixing efficiency of \textit{any} pseudo-marginal MH algorithm. 
 This
bound, combined with an assumption that the noise
in the log-target is Gaussian with a variance that is inversely
proportional to the cost in processing time per iteration leads to a bound on the
overall efficiency of the algorithm in terms of effective samples per
second. The bound on overall efficiency is a function of the noise variance and it was
shown that the variance at which it is optimised lies between $0.85$
and $2.82$, with the exact value depending on the efficiency of the
idealised marginal algorithm. Since this applies to any algorithm it
therefore applies to the PsMRWM across any range of scalings and
implies a degree of insensitivity of the optimal variance to the
choice of scaling. The result presented herein complements that of
\cite{Doucetetal:2015}, and this is discussed further in Section
\ref{sect.discuss}. 

The main theoretical result of this article, Theorem
\ref{thrm.insensitive.mu}, is stated and proved in Section
\ref{sect.main.thrm}. Given that $\int\md w~g(w)e^w$ is finite, $g$
cannot, at least in terms of its tail behaviour, be `too far'
from log-concave. In Section \ref{sect.nonlcc} we demonstrate empirically 
that the statement in Theorem \ref{thrm.insensitive.mu} that relies on the log-concavity
appears to hold more generally.
 The efficiency measure upon which Theorem
\ref{thrm.insensitive.mu} is based relies on several assumptions, in
particular it is a limit result for high dimensional targets and
it relies on the noise in the proposal and the
proposed position in the target being independent.
Section \ref{sect.sim.study} examines two simulation studies for the
insensitivity properties predicted by Theorem
\ref{thrm.insensitive.mu}. Firstly, the simulation study of
\cite{STRR:2014}, where the estimate of the target was obtained from a
particle filter, then a new simulation study
where the estimate of the target is obtained by importance
sampling; both studies support the heuristics of Theorem
\ref{thrm.insensitive.mu}. The article concludes with a discussion.
 
\section{Set-up and main theoretical result}
\label{sect.main.thrm}
\subsection{The efficiency function}
\label{sect.revSTRR}
\cite{STRR:2014} consider a sequence of targets
$\left\{\pi^{(d)}(x^{(d)})\right\}_{d=1}^\infty$. In each dimension, $d$, an
unbiased estimator is available, exactly as described in and around
Equation \eqref{eqn.estimate.prop}. 
It is assumed that there exists a constant, $s^{(d)}$ such that
\[
\lim_{d\rightarrow \infty}\frac{1}{s^{(d)}}\Norm{\nabla \log
  \pi^{(d)}(X^{(d)})}^2
=1
~~~\mbox{and}
~~~
\lim_{d\rightarrow \infty}\frac{1}{s^{(d)}}\nabla^2\log
  \pi^{(d)}(X^{(d)})=-1,
\]
where $X^{(d)}\sim \pi^{(d)}$, and a regularity
condition on the target allows same $s^{(d)}$ to be used in
both expressions. The constant $s^{(d)}$ is a measure of
the roughness of $\log \pi^{(d)}$; for example, if
\begin{equation}
\label{eqn.prod.targ}
\pi^{(d)}(x^{(d)})=\exp\left(\sum_{i=1}^df\left(x_i^{(d)}\right)\right)
\end{equation}
 then
$s^{(d)}=-d/\Expect{f''(X)}$, where $X$ has density $\exp(f(x))$. 
The scaling for the RWM algorithm in dimension $d$ is then set to
\begin{equation}
\lambda^{(d)}=\ell/\sqrt{s^{(d)}},
\end{equation}
for some fixed $\ell$, and the proposal is $X^{*(d)}=x^{(d)}+\lambda^{(d)}Z^{(d)},~Z^{(d)}\sim N(0,I)$.
The Markov chain on $(X,W)$ is assumed to be stationary and 
the distribution of the additive
noise in the proposal is assumed to be independent of the position
\begin{equation}
\label{eqn.indep.assumption}
g(w^*|x^*)=g(w^*).
\end{equation}
This assumption is made for tractability although it has been
found to hold approximately in simuation studies on real statistical examples
\cite[]{STRR:2014,Doucetetal:2015}. 

Perhaps the most natural measure of efficiency of an MCMC algorithm
is the effective sample
size \cite[ESS, e.g.][Ch.3]{CarlinLouis:2009} of each component; the
number of independent samples that would lead to the 
 same variance in the estimator of the posterior mean of
the component as
that arising from the correlated sample of points obtained from
the MCMC algorithm. Even this measure, however, has its drawbacks,
since it is not invariant to a transformation of the target.
\cite{STRR:2014} examine the efficiency of the RWM in terms
of expected squared jumping distance (ESJD) on the sequence of
targets.   
Subject to further technical
conditions on the sequence  
 it is shown that the limiting ESJD has the form
\begin{equation}
\label{eqn.ESJD}
J_m(\ell)=2\ell^2\Expect{\Phi\left(\frac{B}{\ell}-\frac{\ell}{2}\right)}.
\end{equation}
Here $B:=W^*-W$ is the difference in the additive noise in the estimate
of $\log \pi$ at the proposed value and at the current value, and
$\Phi$ denotes the cumulative distribution function of a standard
Gaussian random variable. 
Maximising ESJD is equivalent to minimising the lag-$1$
autocorrelation of the chain. The following result (proved in Appendix
\ref{app.prove.prop.pos}) extends results on the positivity of Metropolis-Hastings
algorithms in Lemma 3.1 of \cite{Baxendale:2005} and Proposition 3 of \cite{Doucetetal:2015} to the
pseudo-marginal RWM. It shows that for jump proposal distributions
such as the Gaussian or Student-t all of the eigenvalues of the
algorithm are non-negative; hence, minimising the lag-$1$
autocorrelation is a sensible goal.
\begin{proposition}
\label{prop.positive}
If the proposal in a Metropolis-Hastings algorithm
satisfies
\begin{equation}
\label{eqn.q.pos}
q(x^*|x)=\int r(x,z)r(x^*,z)\md z,
\end{equation}
then the corresponding pseudo-marginal Metropolis-Hastings algorithm
is positive.
\end{proposition}
 Further justification for the use
of ESJD as a measure of efficiency 
is provided in \cite{STRR:2014} where it is shown
that for the product target in \eqref{eqn.prod.targ}, and subject to
further technical conditions, as 
$d\rightarrow \infty$ a scaled version of the first component of each
element in the sequence of Markov chains converges to a
diffusion, the speed of which is proportional to $J_m(\ell)$. When a
limiting diffusion exists, then in that limit $J_m(\ell)$ is also proportional to the ESS
and is invariant (up to a multiplicative constant) to any differentiable transformation, hence
$J_m(\ell)$ is unambiguously the right measure of efficiency.

\subsection{Insensitivity}
Our main result refers to the situation when
there is no noise in the estimate of $\pi$, $B=0$, when the limiting
ESJD simplifies to 
\begin{equation}
\label{eqn.ESJD.infty}
J_\infty(\ell)=2\ell^2\Phi\left(-\frac{\ell}{2}\right).
\end{equation}
In this case, as noted in \cite{Roberts/Gelman/Gilks:1997}, the optimal scaling is $\lho\approx 2.38$.

When the additive noise in the log-target is Gaussian then \eqref{eqn.ESJD} is particularly
tractable and \cite{STRR:2014} suggest through a plot and an
asymptotic argument that 
$\hat{\ell}$ is between $\lho$ and $2\sqrt{2}$, where the exact value
depends on the variance of the Gaussian distribution. We show this
rigorously, and for a more general form of noise distribution. We also
provide bounds on the potential loss of efficiency suffered by
choosing a different scaling between $\lho$ and $2\sqrt{2}$.

\begin{theorem}
\label{thrm.insensitive.mu}
Let $\hat{\ell}_m$ and $\lho\approx 2.38$ be the values which
optimise the efficiency
functions $J_m(\ell)$ and $J_\infty(\ell)$ that are defined in
\eqref{eqn.ESJD} and \eqref{eqn.ESJD.infty}. Let $g(w^*)$ be the density of $W^*$, the noise in
the log-target at a proposed new target value, and assume that $W^*$ is
independent of that target value. Then
\begin{enumerate}
\item \label{thrm.lo} $\hat{\ell}_m\ge \lho$.
\item \label{thrm.hi} If $g(w^*)$ is log-concave then $\hat{l}_m\le 2\sqrt{2}$.
\item \label{thrm.eff} For any two scalings, $\ell_1$ and $\ell_2$, both in
  $[\lho,2\sqrt{2}]$, $J_m(\ell_1)/J_m(\ell_2)>0.70$.
\end{enumerate}
\end{theorem}

\subsection*{Proof of Theorem \ref{thrm.insensitive.mu}}
\label{sec.proof.insensitive}
For simplicity of notation we suppress the subscript $m$ throughout this proof.
From \eqref{eqn.joint.posterior} and the independence of $W^*$ from
$X^*$, the density of the noise in the 
log-target at the current value, $W$, is $e^{w}g(w)$. Let $B$ have
density $\rho(b)$ and note that
\begin{equation}
\label{eqn.h.def}
h(b):=e^{b/2}\rho(b)=\int_{-\infty}^\infty\md w~g(w)e^{b/2+w}g(w+b)
=\int_{-\infty}^\infty\md w~g(w+b/2)g(w-b/2)e^{w}
\end{equation}
is a symmetric function, $h(b)=h(-b)$.
Define 
\begin{equs}
f(b,\ell):=\ell^2\left[e^{-b/2}\Phi\left(\frac{b}{\ell}-\frac{\ell}{2}\right)
+e^{b/2}\Phi\left(-\frac{b}{\ell}-\frac{\ell}{2}\right)\right].
\label{eqn.f.for.prop}
\end{equs}
Using \eqref{eqn.ESJD} and \eqref{eqn.h.def}, the squared jumping distance is
\begin{eqnarray}
\nonumber
J(\ell)&=&
2\ell^2\int_{-\infty}^{\infty}
\mbox{d}b~\rho(b)~\Phi\left(\frac{b}{\ell}-\frac{\ell}{2}\right)
=
{2\ell^2}\int_{-\infty}^{\infty}
\mbox{d}b~h(b)~e^{-b/2}\Phi\left(\frac{b}{\ell}-\frac{\ell}{2}\right)\\
&=&
{2}\int_{0}^{\infty}
\mbox{d}b~h(b)~f(b,\ell),
\label{eqn.J.formA}
\end{eqnarray}
by the symmetry of $h$.
From \eqref{eqn.ESJD.infty}, straightforward differentiation gives:
\begin{eqnarray}
\label{eqn.J.inf.first}
\frac{d}{d\ell}(\log J_{\infty})&=&\frac{2}{\ell} - \frac{\phi(\ell/2)}{2\Phi(-\ell/2)},\\
\label{eqn.J.inf.second}
\frac{d^2}{d\ell^2}\left(\log J_{\infty}\right)&=&-\frac{2}{\ell^2}-\frac{\phi(\ell/2)}{4\Phi(-\ell/2)^2}\left[\phi(\ell/2)-\frac{\ell}{2}\Phi(-\ell/2)\right]<0~\forall~\ell>0,
\end{eqnarray}
so that (for $\ell>0$) $J_{\infty}$ has a single stationary point (at $\lho$), which is a maximum.

Lemma \ref{res.properties.f} provides key properties of $f$. Its
proof is non-trivial but uninteresting and so is deferred to Appendix \ref{sec.proof.lemma}.

\begin{lemma}
\label{res.properties.f}
For all $b\ge 0$, the following hold.
\begin{enumerate}
\item \label{f.dfdl.bds} 
\[
\frac{2}{\ell}-\frac{\phi(\ell/2)}{2\Phi(-\ell/2)}<
\frac{1}{f}\frac{\partial f}{\partial \ell}
<
\frac{2}{\ell}.
\]
\item \label{f.dfdb.dfdl}
\[
\frac{\partial f}{\partial \ell}
=
\ell \frac{\partial^2f}{\partial b^2} + 
\left(\frac{2}{\ell}-\frac{\ell}{4}\right)f.
\]
\item \label{f.dfdb.lim} 
$\partial f/\partial b\rightarrow 0$ as $b\rightarrow 0$ and as
  $b\rightarrow \infty$, whatever the value of $\ell>0$.
\item \label{f.dfdb}
For all $\ell>0$, $\partial f/\partial b \le 0$.
\end{enumerate}
\end{lemma}

Combining Part \ref{f.dfdl.bds} of Lemma \ref{res.properties.f} with \eqref{eqn.J.inf.first} gives
$f~d\log J_{\infty}/d\ell<\partial f/\partial
\ell<2f/\ell$. Multiplying by $h$, which is non-negative, integrating
and using \eqref{eqn.J.formA} we then obtain
\begin{equation}
\label{eqn.bds.deriv.J}
\frac{d}{d\ell}(\log J_{\infty})<\frac{d}{d\ell}(\log J)<\frac{2}{l}.
\end{equation}

We now proceed with the proof of Theorem \ref{thrm.insensitive.mu}.

\textit{Proof of Part \ref{thrm.lo} of Theorem
  \ref{thrm.insensitive.mu}}: 
by \eqref{eqn.J.inf.second}, for $\ell<\lho$, $dJ_{\infty}/d\ell>0$
and so $d\log J_{\infty}/d\ell>0$. The result then follows from \eqref{eqn.bds.deriv.J}. 

\textit{Proof of Part \ref{thrm.hi} of Theorem \ref{thrm.insensitive.mu}}:
from the definition in \eqref{eqn.h.def}, 
\begin{equation}
\label{eqn.dhdb}
\frac{\partial h}{\partial b}
=\frac{1}{2}\int_{-\infty}^{\infty}\md w~
g(w-b/2)g(w+b/2)e^w\left(\frac{g'(w+b/2)}{g(w+b/2)}
-\frac{g'(w-b/2)}{g(w-b/2)}\right)\le 0\mbox{ for }b\ge 0
\end{equation}
by the log-concavity of $g$.

Furthermore, $\exists~\overline{g}~s.t.~g(w)\le\overline{g}<\infty$ (since $g$ is a log-concave density) and
hence  by \eqref{eqn.h.def}
\begin{eqnarray}
\label{eqn.h.bound.one}
h(0)&\le&
\overline{g}\int_{-\infty}^{\infty}\md w~g(w)e^w=\overline{g},\mbox{ and}\\
\label{eqn.h.bound.two}
h(b)&\le&
\overline{g}\int_{-\infty}^{\infty}\md w~g(w+b)e^{b/2+w}
=\overline{g}e^{-b/2}\int_{-\infty}^{\infty}\md w~g(w)e^{w}=\overline{g}e^{-b/2}.
\end{eqnarray} 

By Part \ref{f.dfdb.dfdl} of Lemma \ref{res.properties.f},
\begin{equation}
\label{eqn.dJdl}
\frac{\mbox{d} J}{\mbox{d} \ell}
=
2\ell\int_0^\infty \mbox{d}b~h(b)\frac{\partial^2f}{\partial b^2}
+ 
\left(\frac{4}{\ell}-\frac{\ell}{2}\right)\int_0^\infty \mbox{d}b~h(b)~f(b,\ell).
\end{equation}
The first term is
\[
2\ell\left[h(b)\frac{\partial f}{\partial b}\right]_0^\infty
-2\ell\int_0^\infty \mbox{d}b~\frac{\partial h}{\partial b}\frac{\partial f}{\partial b}.
\]
Now $\left[h(b)\frac{\partial f}{\partial b}\right]_0^\infty=0$ by
\eqref{eqn.h.bound.one}, \eqref{eqn.h.bound.two} and
Part \ref{f.dfdb.lim} of Lemma \ref{res.properties.f}.
Also $\partial f/\partial b\le 0$  by
Part \ref{f.dfdb} of Lemma \ref{res.properties.f}, and $\partial
h/\partial b \le 0$ by \eqref{eqn.dhdb}; thus the first term in
(\ref{eqn.dJdl}) cannot be positive. 
The second term in \eqref{eqn.dJdl} is guaranteed to be negative provided $\ell>2\sqrt{2}$.

\textit{Proof of Part \ref{thrm.eff} of Theorem \ref{thrm.insensitive.mu}}:
 by \eqref{eqn.J.inf.second},
 for $\ell \in[\lho,2\sqrt{2}]$ (and, indeed, above this), the lower bound in
\eqref{eqn.bds.deriv.J} is always negative; also the upper bound is always
positive. Supposing, without loss of generality, that $\ell_2>\ell_1$,
we therefore have
\[
[\log J_{\infty} ]_{\lho}^{2\sqrt{2}}
\le
[\log J_{\infty}]_{\ell_1}^{\ell_2}
<[\log J]_{\ell_1}^{\ell_2}
<
[2\log \ell]_{\ell_1}^{\ell_2}
\le
[2\log \ell]_{\lho}^{2\sqrt{2}}.
\]
Evaluating the outer-most terms and exponentiating gives (to 3dp)
\[
 0.949~J(\ell_1)<J(\ell_2)< 1.411~J(\ell_1).
\]

\section{The log-concavity condition}
\label{sect.nonlcc}
The lower bound for $\hat{\ell}$ in Theorem \ref{thrm.insensitive.mu} holds for all noise
distributions whereas the upper bound has only been shown to hold when
$W^*$ has a log-concave density. This condition is weaker than
might be thought, holding, for example, when the unbiased multiplicative
noise, $e^{W^*}$, has a (left-truncated) $t$
distribution or a Gamma distribution, even if the Gamma shape parameter is
less than unity. Nonetheless 
it is natural to ask whether or not the upper bound holds more
generally. The key consequence of the log-concavity of $g_*$ is that
$\partial h/\partial b \le 0$ for $b\ge 0$. However it is clear from
the proof that a weaker (yet still sufficient) condition for the upper bound is
\[
\int_0^{\infty}\md b~ \frac{\partial h}{\partial b}\frac{\partial
  f}{\partial b}
>0.
\]
Clearly there is scope for $\partial h/\partial b>0$ over
some regions whilst the whole expression in \eqref{eqn.dJdl} remains negative,
so log-concavity is certainly not a necessary condition. 

We investigate the following set of discrete noise distributions, indexed by
$p\in (0,1)$ and $\epsilon \in (0,1)$:
\[
e^{W^*} = 
\left\{
\begin{array}{lll}
\epsilon&w.p.&p^*\\
a&w.p.&1-p^*.
\end{array}
\right.,
\]
where $a=(1-p^*\epsilon)/(1-p^*)$.
In this case
\[
J_{\epsilon,p^*}=2\ell^2\left[p^*(1-p)\Phi(-k/\ell-\ell/2)+(p^*p+(1-p^*)(1-p))\Phi(-\ell/2)+(1-p^*)p\Phi(k/\ell-\ell/2)\right],
\]
where $k=\log a-\log \epsilon$, and $p=p^*\epsilon$.

The top-left panel of Figure \ref{fig.BGP} shows the optimal scaling as a function of the two
noise parameters $\hat{\ell}(\epsilon,p^*)$ and demonstrates that for
this set of noise distributions $2.38< \hat{\ell}< 2.64$. 
Indeed, we have not been able to find a model for $W^*$ where
$\hat{\ell}>2\sqrt{2}$ and we conjecture that $\hat{\ell}\le 2\sqrt{2}$
whatever the distribution of $W$.

\section{Simulation study}
\label{sect.sim.study}
We first briefly discuss theoretical results that are available for
likelihoods that are estimated via particle filters and via importance
sampling. We then describe the evidence of insensitivity arising from
the simulation study in \cite{STRR:2014}, which used a particle
filter, before describing a new simulation study that uses importance sampling.

When a particle filter with $m$ particles is used to estimate a likelihood for a process
observed over large number of time points it is to be expected
\cite[]{berard2013lognormal} that the distribution of repeated
estimates of the log-target will be approximately Gaussian with a
variance, $\sigma^2\propto 1/m$. \cite{STRR:2014} found this be the case for  $m\ge 100$, although with
$m=50$, the variance was considerably larger than expected and
the distribution of estimates had a heavier left tail and a lighter
right tail than the corresponding Gaussian. In Lemma 2 of \cite{Pittetal:2012} 
the delta method is used to show
that even when the likelihood is estimated via importance
sampling, in the limit as $m\rightarrow \infty$ the log-likelihood will also
be Gaussian with a variance $\sigma^2\propto 1/m$; an optimal variance of about $0.92^2$ is also suggested. However, straightforward examination of the error terms
shows that the delta method requires $\sigma^2<<1$, so an overall variance
of $0.92^2$ would only be achievable if the log-likelihood were the
sum of a number of terms, each of which could be estimated separately.
Our new simulation study will deliberately consider an example where
this is not the case.

\cite{STRR:2014} examined the five-dimensional target distribution that
arises from a continuous-time Markov jump process (the Lotka-Volterra
predator-prey model), noisy observations of which are available at a
set of $50$ time points. 
A pilot run provided an estimate of the posterior variance matrix, $\Vhat$, for
the five parameters, and the jump proposal was  as in \eqref{eqn.RWM.prop.prac}.

Since $\ell \propto \lambda$, with the constant of
proportionality unknown for any real target, to test Parts 1 and 2
of Theorem \ref{thrm.insensitive.mu} we must consider the
ratio of upper and lower end points and compare against $2\sqrt{2}/\lho\approx 1.19$. There is considerable Monte Carlo
variability in the efficiencies displayed in Figure 6 in \cite{STRR:2014}; nonetheless, over the large range of $m$ values
considered, the largest optimal scaling was no more than twice the
smallest optimal scaling. It is also clear from the same figure that over the range of
optimal scalings, for each $m$ the efficiency over this range is at
least $70\%$ of the maximum. Finally, the insensitivity result of
\cite{Doucetetal:2015} is also supported as the optimal variance
(estimated at the posterior mean for $x$) ranges between $0.97$ and $2.15$. 

The simulation study of \cite{STRR:2014} had $d=5$ and an additive noise
distribution that was close to Gaussian and with a variance that was inversely proportional
to the computational cost.
The theory in \cite{STRR:2014} is strictly valid in the limit as
$d\rightarrow \infty$, yet even with this low dimension there is
evidence that the optimal scaling was relatively
insensitive to the choice of $m$. The range of variation  was not as
narrow as predicted by Theorem \ref{thrm.insensitive.mu}, although some of the excess
could have been due to Monte Carlo error. 

We wish to investigate the applicability of Theorem
\ref{thrm.insensitive.mu} further. We therefore 
conduct a simulation study 
based on a real statistical model but using
importance sampling rather than a particle filter so that the additive
noise is not expected to be Gaussian (nor, indeed, is its variance expected to
be inversely proportional to the computational cost).

\subsection{Logistic regression using a latent Gaussian process}
\label{sect.GP}
\cite{FilipponeGirolami:2014} use pseudo-marginal Metropolis-Hastings
to obtain the posterior distribution of the parameters of a latent Gaussian
process (GP) where the observed response is Bernoulli with a success
probability determined from the GP via the probit link function.  \cite{GiorgiDiggle:2015} use
Monte Carlo maximum likelihood to estimate the parameters of a
generalised linear geostatistical model for binomial data where the
success probability depends on a latent GP and on fixed
effects via the logistic link function. In both of the above articles 
the likelihood for a particular set of parameter values 
is estimated using importance
sampling with the proposal based upon the Laplace approximation
or the Expectation Propagation algorithm
\cite[]{FilipponeGirolami:2014} or a variation on the Laplace
approximation \cite[]{GiorgiDiggle:2015}. 
Our statistical model is motivated by these two
applications. 

Let
$z_i~(i=1,\dots,l)$ be a set of points in $\mathbb{R}^a$ with
components $z_{ik},~(k=1,\dots a)$ and let $Z$ be the $l\times a$
matrix with $i$th row $z_i'$. We use the logistic link
function and denote the overall mean on the logit scale by
$\mu\in \mathbb{R}$ and covariate effects by $\beta\in \mathbb{R}^a$.
 The variance of the GP is $\tau^2\in \mathbb{R}^+$ and the range
 parameters (one for each dimension of the process) are 
 $\phi\in\left(\mathbb{R}^+\right)^a$, so that the correlation between
 the values of the GP at the $l$ points is the $l\times l$ matrix $R$
 with elements 
\[
R_{ij}=\exp\left(-\sqrt{\sum_{k=1}^a\left(\frac{z_{ik}-z_{jk}}{\phi_k}\right)^2}\right).
\]
We consider the following statistical model:
\begin{eqnarray*}
S|\phi,\tau^2&\sim&N_{l}(0,\tau^2R)\\
p_i&=&\frac{\exp(s_i+\mu+z_i^t\beta)}{1+\exp(s_i+\mu+z_i^t\beta)}\\
Y_i|s_i,\mu,\beta&\sim& \mbox{Bin}(n,p_i).
\end{eqnarray*}
Since all of the importance sampling
algorithms in \cite[]{FilipponeGirolami:2014} and \cite[]{GiorgiDiggle:2015} 
require an iterative scheme to obtain the proposal
distribution, we opt instead for a simpler approach based on ideas
for Poisson data in \cite{HaranTierney:2012} and
\cite{Lampaki:2015}. We first transform the data as follows:
\[
y^+_i=
\left\{
\begin{array}{ll}
1/2&\mbox{if}~y_i=0,\\
n-1/2&\mbox{if}~y_i=n,\\
y_i&\mbox{otherwise}
\end{array}
\right.
~~~,~~~y_i^*=\mbox{logit}\left(\frac{y_i^+}{n}\right).
\]
Using the delta method, the expectation and variance of $Y^*_i$ given
the GP are
respectively
\[
\Expect{Y^*_i|s_i}\approx s_i+\mu+z_i'\beta~~~\mbox{and}~~~\Var{Y^*_i|s_i}\approx\frac{1}{np_i(1-p_i)}. 
\]
For tractability we approximate $p_i$ in the variance term using
the observed data: $p_i\approx y_i^+/n$. This leads to a Gaussian
approximation of
\[
Y^*|s\sim N_l\left(s+\mu\bmone+Z\beta,D\right),
\]
where $D$ is a diagonal matrix with $1/D_{ii}=y^+_i(1-y^+_i/n)$, and
$\bmone$ is an $l$-vector of ones. Combining of this with the
Gaussian prior for $S$ leads to a Gaussian approximation for
$S|y^*$ with mean $\mu_c$ and variance $\Sigma_c$, obtained via
standard formulae. The proposal distribution for our importance
sampler is a
Student-t distribution with $\nu=20$ degrees of freedom and density
\[
q(s|y)\propto \left(1+\frac{1}{\nu}(s-\mu_c)'\Sigma_c^{-1}(s-\mu_c)\right)^{-\frac{\nu+\ell}{2}}.
\]

We consider $a=4$ so that $d=10$, and apply the following map:
\[
(\mu,\beta_1,\dots,\beta_4,\log(\tau^2),\log(\phi_1),\dots,\log(\phi_4))\leftrightarrow x.
\]
We place $81$ points, $z_i$, uniformly on a
hypergrid with oppposite corners at
$(-\frac{1}{2},-\frac{1}{2},-\frac{1}{2},-\frac{1}{2})$ and
$(\frac{1}{2},\frac{1}{2},\frac{1}{2},\frac{1}{2})$. A data set was
simulated using $n=10$ and $x=(\frac{1}{2},-1,0,0,1,0,0,0,0,0)$. For the
analysis we assume \textit{a priori} $X\sim N_{10}(0,I)$; this
prior is tight enough to prohibit difficult tail behaviour (the
investigation of which is not the point of this simulation study), yet
relaxed enough that the main influence is due to the likelihood (the
mean diagonal term of the posterior variance matrix was $0.35$, and
none of the terms was larger than $0.5$). 

Define the sets of possible scalings, $\Lambda$, and number of
importance samples, $\mathcal{M}$ as
\[\Lambda:=\{0.2,0.4,0.6,0.7,0.8,1.0,1.2,1.4,1.6\},~\mbox{and}~
 \mathcal{M}:=\{10,20,40,100,200,400,1000\}.
\]
The posterior variance
matrix, $\hat{V}$, was estimated from a
trial run and for each combination of
$\lambda\in\Lambda$ and $m\in\mathcal{M}$, 
a pseudo marginal RWM was run using the proposal in
\eqref{eqn.RWM.prop.prac}. At least $2\times 10^5$ iterations were
used, with the number increasing as $m$ decreased so as to ensure that
the effective sample size of any component was always greater than
$1000$. 

For each $m\in\mathcal{M}$ and $\lambda\in\Lambda$ 
define the relative efficiencies over $\lambda$ and over $m$,
respectively as
\[
\ESSrl_{m,\lambda}:=
\frac{\ESS_{m,\lambda}/T_{m,\lambda}}{\max_{\lambda\in\Lambda}(\ESS_{m,\lambda}/T_{m,\lambda})}
~~~\mbox{and}~~~
\ESSrm_{m,\lambda}:=
\frac{\ESS_{m,\lambda}/T_{m,\lambda}}{\max_{m\in\mathcal{M}}(\ESS_{m,\lambda}/T_{m,\lambda})},
\]
where $\ESS_{m,\lambda}$ is the minimum effective sample size over the
$d=10$ components of $x$, and $T_{m,\lambda}$ is the CPU time for the run.

\begin{figure}
\begin{center}
\subfigure{
  \includegraphics[scale=0.45,angle=0]{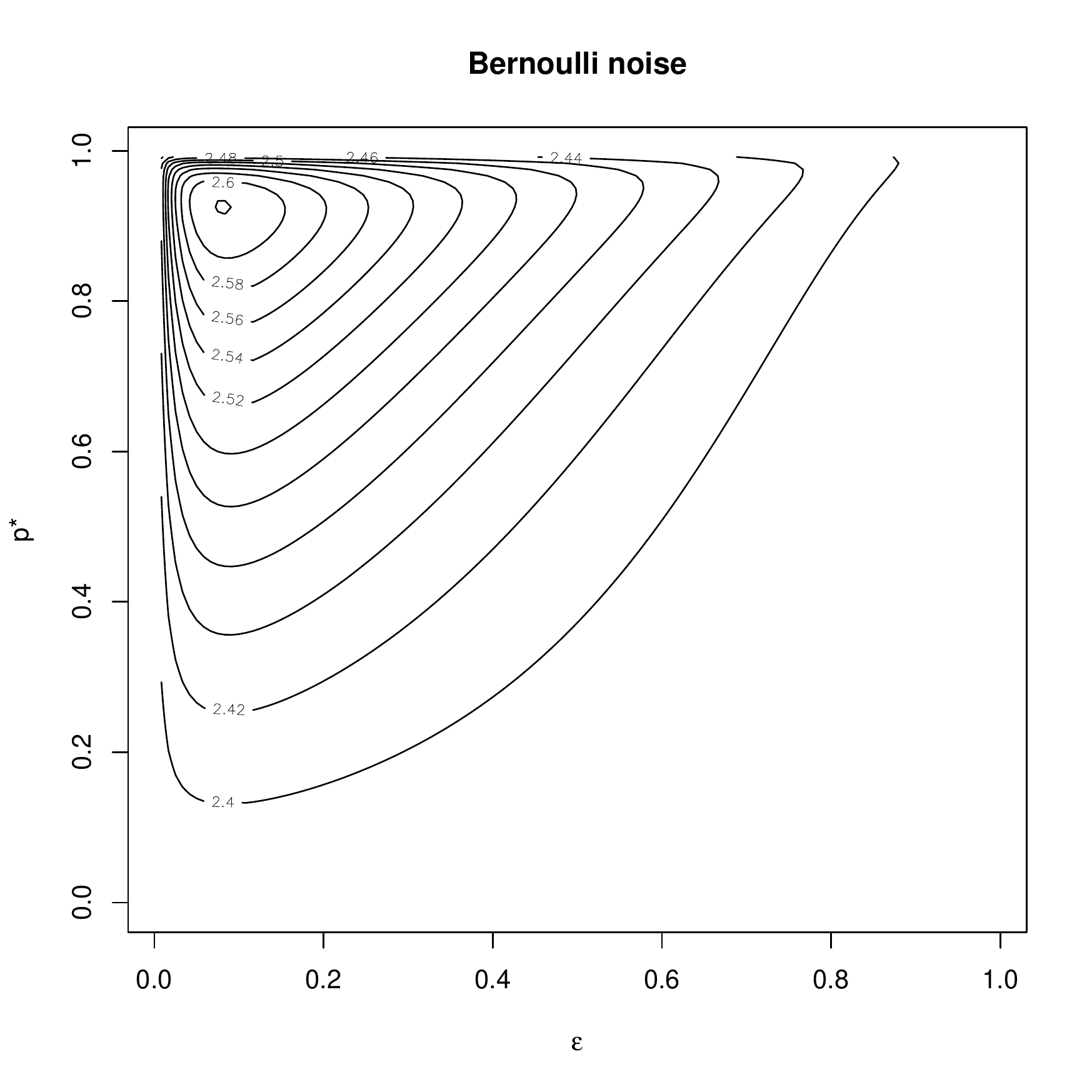}
  \includegraphics[scale=0.45,angle=0]{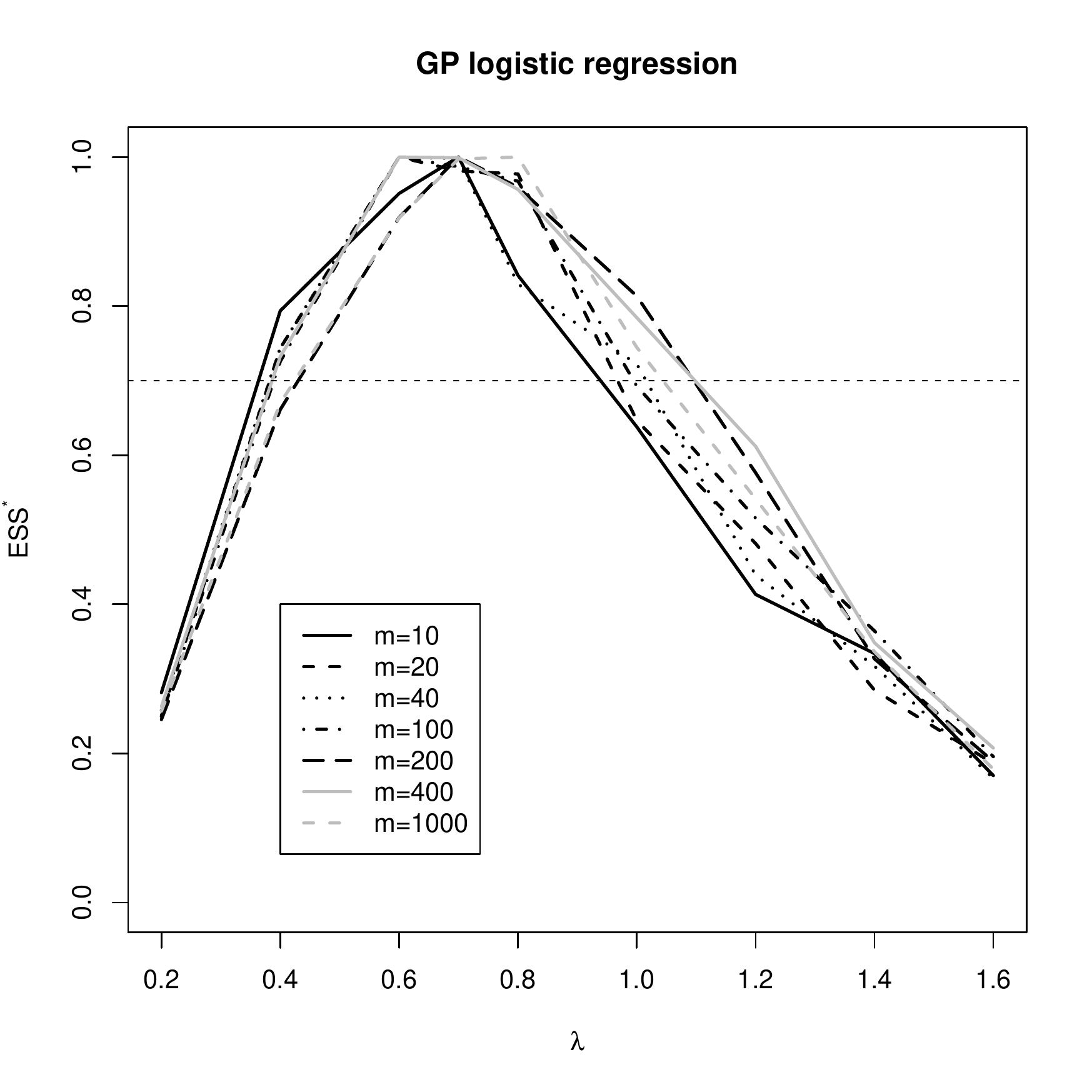}
}
\subfigure{
  \includegraphics[scale=0.45,angle=0]{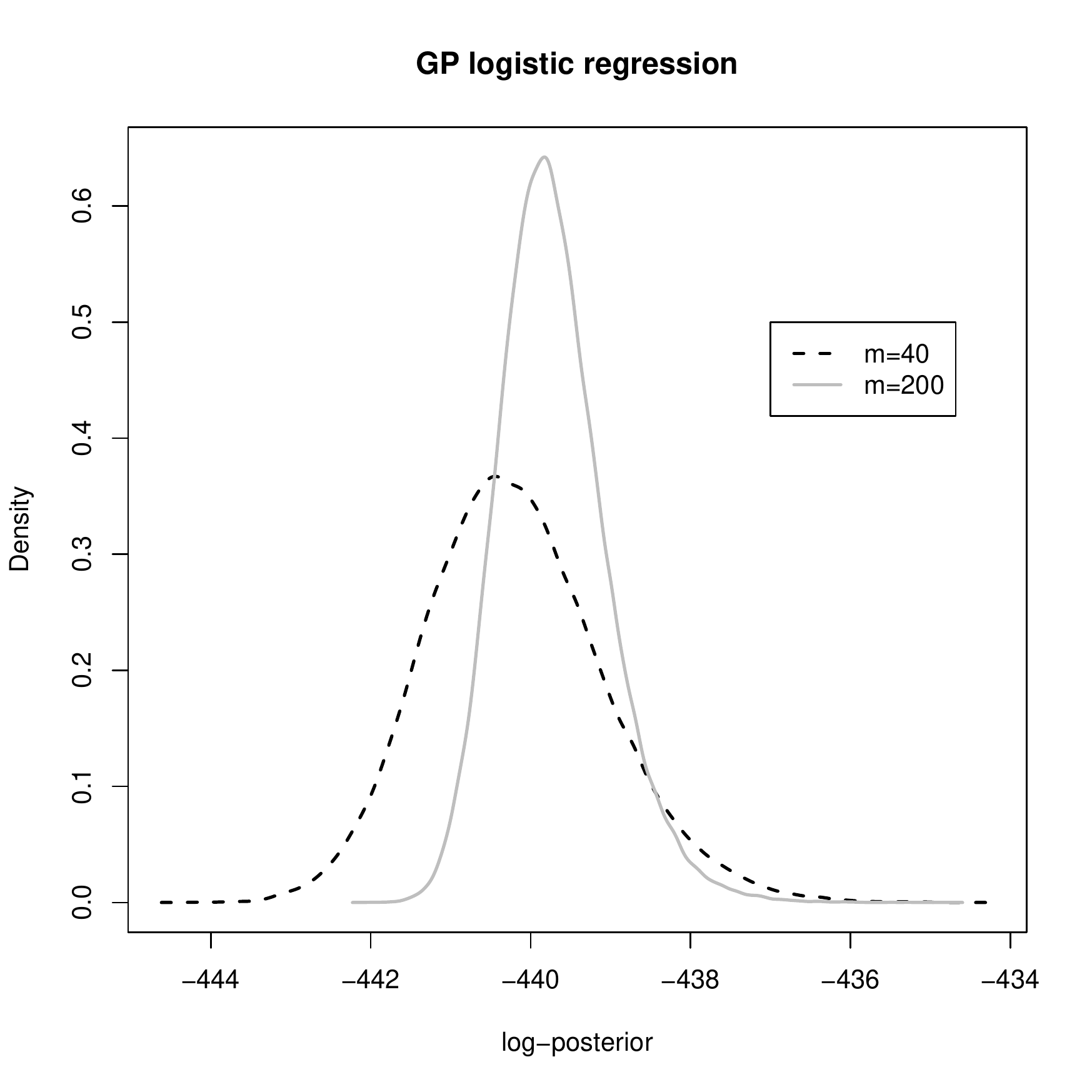}
  \includegraphics[scale=0.45,angle=0]{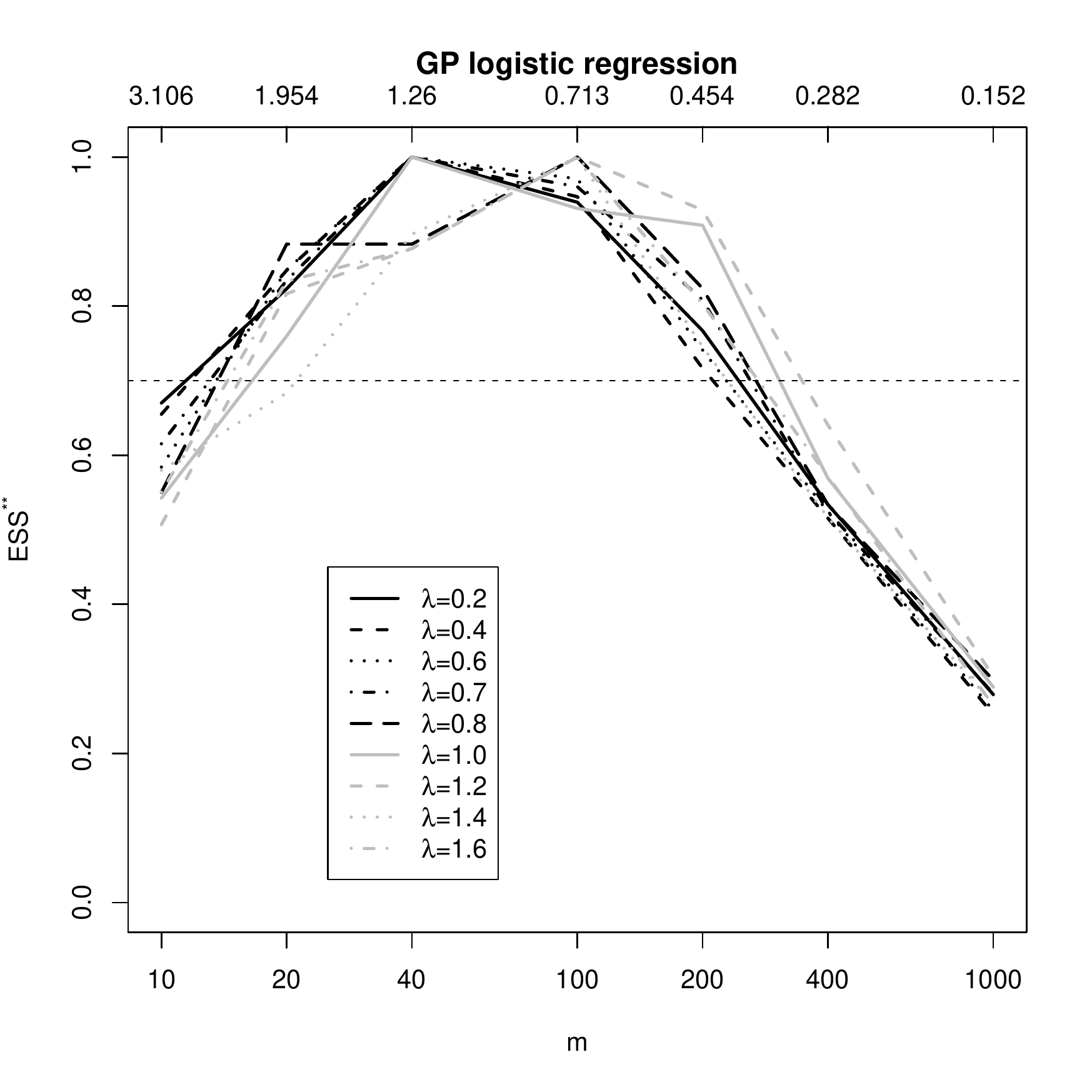}
}
\caption{
Top-left panel: optimal scaling for the Bernoulli noise model as a function of the two
noise parameters, $\epsilon$ and $p^*$. Remaining panels: results from the Gaussian Process regression; top-right: $ESS^*$
against scaling for each $m\in \mathcal{M}$; bottom-left: kernel
density estimate of the distribution of the noise in the log-posterior
(at the posterior mean) for $m=40$ and $m=200$ using $10^5$ samples;
bottom-right: $ESS^{**}$ against $m$ (bottom axis) and variance of the
additive noise (top
axis) for each value of $\lambda\in\Lambda$.
\label{fig.BGP}
}
\end{center}
\end{figure}

The top-right panel in Figure \ref{fig.BGP} shows, for each $m\in \mathcal{M}$,
a plot of $\ESSrl_{m,\lambda}$ against $\lambda$. For each $m$, the
optimal scaling always lies in the narrow range between $0.6$ and
$0.8$. 
Furthermore,
 the efficiency is always at least $70\%$ of the
optimal obtainable efficiency over a much wider range than this,
approximately between $0.4$ and $1.0$. This provides evidence that
 the insensitivity and
robustness predicted by Theorem \ref{thrm.insensitive.mu} can continue hold for moderate
dimensions and when the target is not estimated using a particle filter.

The bottom left panel in Figure \ref{fig.BGP} shows kernel density plots of the
estimated log-posterior at the posterior mean for $x$, when $m=20$ and $m=200$,
two values that bound the range of sensible values for $m$ for this
problem (see the discussion of the third panel, below). Unlike the
discrepancy from a Gaussian distribution that was found in the
particle filter example in
\cite{STRR:2014} (and indeed in the particle filter example in
\cite{Doucetetal:2015} with $m=4$) it is the
right tail that is too heavy and the left tail that is too light
(skewness=$0.22$ and $0.62$ respectively), and
this persists across the range of useful $m$ values. To guage the
variability of the variance and skewness across the posterior for one
of the most efficient $m$ values, $1000$
independent samples of $x$ from the posterior were obtained by
thinning a run of $10^6$ iterations, which had a minimum ESS of
$12~634$, by a factor of $1000$. For each $x$ value, the log-target
was estimated a thousand times using $m=100$, and the variance and
skewness were noted. The $(0.025,0.5,0.975)$ quantiles for the
variance and skewness were, respectively, $(0.54,0.74,0.96)$ and
$(0.32,0.52,0.75)$, showing a moderate amount of variability over the
main posterior mass.  

The bottom right panel in Figure \ref{fig.BGP} shows, for each $\lambda\in\Lambda$,
a plot of $\ESSrm_{m,\lambda}$ against $m$. For each scaling, the optimal
value of $m$ lies between $40$ and $100$, corresponding to variances
of approximately $1.26$ or $0.71$ respectively. Interestingly, also,
the efficiency is around $70\%$ or higher for all $m$ between $20$ and
$200$. This provides evidence that the insensitivity predicted in
\cite{Doucetetal:2015} can continue to hold even when the target is
moderately skewed and, as is clear from the parallel scales for $m$
and the variance of the additive noise, $\sigma^2$, that the variance is \textit{not} inversely proportional to
$m$ (indeed a log-log plot and a simple linear regression show that, approximately
$\sigma^2\propto m^{-0.65}$).

\FloatBarrier

\section{Discussion}
\label{sect.discuss}
The thrust of this article is that the optimal scaling of a
pseudo-marginal RWM algorithm is insensitive to the noise
distribution, and hence, when the noise is generated by
an importance sampler or a particle filter, it is insensitive to the
number of samples or particles, $m$. Moreover, for a particular $m$,
 the loss in efficiency
over the range of optimal scalings, compared with the optimal
efficiency for that $m$ is small.

Theorem \ref{thrm.insensitive.mu} is limited to the pseudo-marginal RWM and is
strictly only proved in the limiting regime of \cite{STRR:2014} which
specifies, in particular, that the distribution of the additive noise
in the proposal should be independent of the proposed position. However
Theorem \ref{thrm.insensitive.mu} requires only the mild log-concavity
assumptions on the form of the noise
distribution. There is an implicit assumption that, for any fixed
noise generating mechanism (e.g. choice of $m$), the computational cost of the
algorithm does not depend on the scaling. This is certainly true for
the example considered in Section \ref{sect.GP} (and similarly in 
\cite{FilipponeGirolami:2014})
and for many other examples
such as inference for partially observed stochastic differential equations using a
particle filter \cite[e.g.][]{GolightlyWilkinson:2011}; however it is unlikely to hold in
other scenarios such as inference for a Markov jump process
\cite[e.g.][]{GolightlyWilkinson:2011,STRR:2014}, where doubling all of the rate parameters effectively
doubles the CPU time required for simulations. Even in this scenario,
however, the dependence on scaling of the total CPU time for a run will be small provided
the scaling is much smaller than the width of the main posterior mass, as happens in
moderate to high dimensions. This is because the average CPU time per
iteration is an average of the costs over a smoothed version of $\pi$:
\[
\int \pi(x)q(x^*|x;\lambda)c(x^*) \md x \md x^*,
\]
where $c(x^*)$ is the computational cost of estimating the target
at $x^*$.

A simulation study in the
literature \cite[]{STRR:2014} with $d=5$ and where the likelihood was estimated
using a particle filter showed the optimal scaling to exhibit an
insensitivity  to the number of particles similar to, though weaker
than, that predicted. For each $m$ value, the CPU time varied with
$\lambda$ by less
than $1\%$ from its mean value and no trend was evident (personal
communication), suggesting that the mechanism discussed above played
no role in the larger-than-expected variability; we conjecture that Monte Carlo variability is at least
partly responsible. A
 new simulation study in this article chose $d=10$ and used importance
sampling to estimate the likelihood; here both the variance and
skewness of the distribution of the
additive noise were shown to vary by a factor of approximately $2$ over the
main posterior mass, yet the insensitivity of the optimal scaling to
the number of importance samples was striking.

\cite{Doucetetal:2015}, in some sense, show a converse result to
Theorem \ref{thrm.insensitive.mu},
that the optimal choice of
$m$ is insensitive to the MCMC algorithm and hence, for an RWM
algorithm, to the choice of scaling. 
The example function $f(x,y)=-(x-y)^2-9\times(x-1)^2$ demonstrates
that neither insensitivity need imply the other
and so the result presented herein and that in \cite{Doucetetal:2015}
complement each other. 
As with \cite{STRR:2014}, the result in \cite{Doucetetal:2015} 
assumes that the distribution of the additive noise
in the proposal is independent of the proposed position. Subject to
this and to the tightness of the upper bound it is valid across all
Metropolis-Hastings algorithms,
and in any dimension, but it, or indeed any future result on insensitivity of the optimal $m$, requires specific assumptions on the form of the noise and its
cost. The key assumptions used in \cite{Doucetetal:2015}, that the
additive noise is Gaussian with a variance inversely proportional to
the computational effort required to obtain it, 
are expected to be valid in the common
scenario where a particle filter is applied to a large number of
observations \cite[]{berard2013lognormal}, however it is
unclear what forms would apply in other situations, and even less clear
how, in a real statistical example, the computational cost, or $m$,
would relate to the parameters of the noise distribution. It is interesting that in our simulation study the
additive noise and the relationship between variance and computational
cost do not satisfy the assumptions of \cite{Doucetetal:2015} and yet the insensitivity of the optimal choice of
$m$ to $\lambda$ still appears to hold.

\appendix
\section{Proof of Proposition \ref{prop.positive}}
\label{app.prove.prop.pos}
Following \eqref{eqn.joint.posterior}, define the
extended target as $\pitil(x,w):=\pi(x)g(w|x)e^w$ and let
\[
c:=\int \md x\md w ~\pitil(x,w)[1-\overline{\alpha}(x,w)]f(x,w)^2\ge
0,
\] 
where 
$\overline{\alpha}(x,w)$ is the average acceptance probability from $(x,w)$.

As in \cite{Baxendale:2005}, note that for $a\ge 0$ and $b\ge 0$,
$a\wedge b=\int_0^\infty\md t~\ind_{[0,a]}(t)\ind_{[0,b]}(t)$. 
Denoting the pseudo-marginal MH
kernel by $P(x,w;x^*,w^*)$, for any $f\in {L}^2(\pitil)$ we have
\[
\begin{array}{lll}
\int \md x\md w\md x^* \md w^* ~\pitil(x,w) P(x,w;x^*,w^*)
f(x,w)f(x^*,w^*)\\
\quad \quad\quad =
c+\int \md x\md w \md x^* \md
w^*~g(w|x)q(x^*|x)g(w^*|x^*)\left[e^w\pi(x)\wedge
  e^{w^*}\pi(x^*)\right]f(x,w)f(x^*,w^*)\\
 \quad\quad \quad=c+\int \md z \int_0^\infty \md t~b(t,z)^2\ge 0, 
\end{array}
\]
where 
\begin{eqnarray*}
b(t,z)&:=&\int \md x \md w ~ g(w|x)f(x,w)r(z,x)\ind_{[0,v\pi(x)]}(t).
\end{eqnarray*}

\section{Proof of Lemma \ref{res.properties.f}}
\label{sec.proof.lemma}
\begin{proof}
Differentiation from the definition of $f$ in \eqref{eqn.f.for.prop} shows that
\begin{eqnarray}
\label{eqn.dfdl}
\frac{\partial f}{\partial \ell}&=&\frac{2}{\ell}f-\ell^2
\phi(\ell/2)e^{-{b^2}/({2\ell^2})},\\
\label{eqn.dfdb}
\frac{\partial f}{\partial b}&=&
\frac{1}{2}\ell^2\left[e^{b/2}\Phi(-b/\ell-\ell/2)-e^{-b/2}\Phi(b/\ell-\ell/2)\right].
\end{eqnarray}

We also note that
\begin{eqnarray}
\nonumber
e^{b/2}\Phi\left(-\frac{b}{\ell}-\frac{\ell}{2}\right)
&=&e^{b/2}\frac{1}{\sqrt{2\pi}}\int_{-\infty}^{-b/\ell-\ell/2}~\mbox{d}t~e^{-t^2/2}\\
&=&\phi\left(\frac{\ell}{2}\right)e^{-{b^2}/{(2\ell^2)}}
\int_{0}^{\infty}~\mbox{d}u~e^{-
u^2/2-u\ell/2}\times e^{-ub/\ell},
\label{eqn.bita}
\end{eqnarray}
and similarly
\begin{equation}
e^{-b/2}\Phi\left(\frac{b}{\ell}-\frac{\ell}{2}\right)
=
\phi\left(\frac{\ell}{2}\right)e^{-{b^2}/{(2\ell^2)}}
\int_{0}^{\infty}~\mbox{d}u~e^{-
u^2/2-u\ell/2}\times e^{ub/\ell}.
\label{eqn.bitb}
\end{equation}
\textit{Proof of Part \ref{f.dfdl.bds}}: combining \eqref{eqn.bita} and \eqref{eqn.bitb} gives
\[
f(b,\ell)=2\ell^2\phi\left(\frac{\ell}{2}\right)e^{-{b^2}/{(2\ell^2)}}
\int_{0}^{\infty}~\mbox{d}u~e^{-
u^2/2-u\ell/2}\times \cosh\left(ub/\ell\right).
\]
Thus,
$f(b,\ell)=2\ell^2\phi\left(\frac{\ell}{2}\right)e^{-{b^2}/{(2\ell^2)}}\times
I(b,\ell)$, where
\[
I(b,\ell)\ge \int_{0}^{\infty}~\mbox{d}u~e^{-
u^2/2-u\ell/2}
=\frac{\Phi(-\ell/2)}{\phi(\ell/2)}.
\]
The result follows on dividing through by $f$ in \eqref{eqn.dfdl}
and applying the above inequality.

\textit{Proof of Part \ref{f.dfdb.dfdl}}: combine \eqref{eqn.f.for.prop},
\eqref{eqn.dfdl} and the fact that
\[
\frac{\partial^2f}{\partial b^2}=\frac{1}{4}f-\ell
\phi(\ell/2)e^{-b^2/(2\ell^2)}.
\]
\textit{Proof of Part \ref{f.dfdb.lim}}: this follows directly from \eqref{eqn.dfdb}.

\textit{Proof of Part \ref{f.dfdb}}: combining \eqref{eqn.bita} and \eqref{eqn.bitb} gives
\[
e^{b/2}\Phi\left(\frac{b}{\ell}-\frac{\ell}{2}\right)
- 
e^{-b/2}\Phi\left(-\frac{b}{\ell}-\frac{\ell}{2}\right)
=
\phi\left(\frac{\ell}{2}\right)e^{-{b^2}/{(2\ell^2)}}
\int_0^\infty\md u~
e^{-
u^2/2-u\ell/2}\times \left(e^{-ub/\ell}
-
 e^{ub/\ell}\right)
<0
\]
since the integrand is negative. The result then follows from this and
\eqref{eqn.dfdb}.

\end{proof}

\bibliographystyle{royal}
\bibliography{biblio}

\end{document}